\documentclass[traditabstract]{aa} 
                                   
\usepackage{graphicx}
\usepackage{amsmath}
\usepackage{txfonts}
\usepackage{natbib}

\bibpunct{(}{)}{;}{a}{}{,} 

\begin{document}
   \title{Preplanetary scavengers: Growing tall in dust collisions}
   \author{Thorsten Meisner
          \inst{1} 
          \and
          Gerhard Wurm
          \inst{1}
          \and
          Jens Teiser
          \inst{1}
          \and
          Mathias Schywek\inst{1}
          }

   \institute{Faculty of physics, University of Duisburg-Essen,
              Lotharstr. 1, D-47057 Duisburg\\
              \email{thorsten.meisner@uni-due.de}
             }

   \date{Received ; accepted }
   
 
  \abstract{
Dust collisions in protoplanetary disks are one means to grow planetesimals, but
the destructive or constructive nature of high speed collisions is still unsettled. In laboratory experiments, we study the self-consistent evolution of a target upon \textit{continuous} impacts of submm dust aggregates at collision velocities of up to 71\,m/s. 
Earlier studies analyzed \textit{individual} collisions, which were more speculative
for high velocities and low projectile masses. Here, we confirm earlier findings that high speed collisions result in mass gain of the target. We also quantify the accretion efficiency for the used $\rm SiO_2$ (quartz) dust sample. For two different average masses of dust aggregates (0.29\,$\mu$g and 2.67\,$\mu$g) accretion efficiencies are decreasing with velocity from 58\% to 18\% and from 25\% to 7\% at 27\,m/s to 71\,m/s, respectively. The accretion efficiency decreases approximately as logarithmic with impact energy. At the impact velocity of 49\,m/s the target acquires a volume filling factor of 38\%. These data extend earlier work that pointed to the filling factor leveling off at 8 m/s to a value of 33\%. Our results imply that high speed collisions are an important mode of particle evolution. It especially allows existing large bodies to grow further by scavenging smaller aggregates with high efficiency.}  



   \maketitle

\section{Introduction}
Particle evolution in protoplanetary disks has seen significant progress over the last few
decades with a number of turns in different directions. 

Astronomical observations of disks in the
visible and near-infrared provide information of dust properties at the surface of 
disks, which shows amorphous or crystalline dust particles of micron size
\citep{vanBoekel2005, olofsson2009}. Aggregation, the change of particle size by sticking collisions, is visible in such observations if submicron grains are assumed to be 
the building blocks. 
These aggregates are still small with respect to planet formation and
it is important to note that small particles are observed over the whole lifetime of 
protoplanetary disks of a few million years. To understand this it is necessary to study collisional evolution. It is not necessarily only ``positive'' aggregation, but it is also the destruction of larger bodies in collisions that might provide the small grains \citep{wurm2005, TeiserWurm2009b, beitz2011, Schraepler2011}.

The experiments in this paper also clearly have this destructive element and in fact, \citet{Dominik2005} showed that aggregation without fragmentation would be so efficient that no small grains would remain observable after a short time. Besides collisional fragmentation other destructive mechanisms are thinkable.
These include gas drag (erosion by wind) as discussed by \citet{paraskov2006} or particle erosion by stellar insolation \citep{Wurm2006b, kelling2009, debeule2013, kocifaj2010}. It has also been suggested that the effect of aggregation is diminished by electrical charging and repulsion of aggregates \citep{Okuzumi2009}.

In any case, the initial mode of particle evolution likely starts from submicron or micron size particles and is the aggregation by hit-and-stick collisions. Much work has been carried out in this field.
As far as astrophysical applications are considered, aggregation has been studied numerically (see, e.g., \citet{Ossenkopf1993, DominikTielens1997, PaszunDominik2009, WadaEtal2009} or \citet{Suyama2008}, and this list is far from complete). It has also been studied experimentally. Cluster-Cluster aggregation by Brownian motion has been found in microgravity experiments \citep{BlumWurm2000, krause2011} and further experiments have studied aggregation of such aggregates \citep{Wurm1998, BlumWurm2000}. The growth of aggregates in a turbulent cloud still marks special regions of space as enigmatic regarding the formation of small aggregates. The core shine in interstellar clouds discovered recently might be 
one example (Steinacker2010).

Within cold environments especially with water ice, a most abundant solid, sublimation, condensation, and sintering are other mechanisms to be considered that can change the initial aggregates \citep{Saito2011, Sirono2006, Aumatell2011, RosJohansen2013, TanakaEtal2013}. High temperature equivalents of sintering might be found very close to a star \citep{Sirono2011, Poppe2010}. The presence of a granular medium or more solid impactors, for example, sintered together, might change the overall picture as well \citep{Colwell2008}.

Besides the observable small particle size scale, the next size steps are currently of critical importance in understanding planet formation.  After initial fractal growth, the compaction of dust aggregates follows as energies in collisions get large enough that particles can restructure. This also occurs in 
numerical simulations as well as experiments \citep{Meakin1988, DominikTielens1997, WadaEtal2011, BlumWurm2000}. Depending on the details of particle size and contact physics, from mm or cm upward, particles are no longer fractal but compact with a porosity or volume filling factor subject to further evolution. This is also a subject of this paper.

Volume filling factor and porosity are complementary, but describe the same thing and are both used randomly thoughout the literature. We note this here to prevent confusion. The volume filling factor $FF$ is defined as 

\begin{equation}
FF = \frac{V_{solid}}{V_{total}},
\end{equation}

where $V_{solid}$ is the volume in an aggregate covered by solids and $V_{total}$
is the total volume of the aggregate. Porosity refers to the void space instead of
the solid space covered and therefore is $1-FF$. To connect to earlier work, we
will use the filling factor when referring to quantitative values here.
When initial restructuring occurs aggregates can have very low filling factors
of $FF << 0.1$ \citep{Kataoka2013}. The details of increasing the filling factor in collisions is still under discussion. While \citet{langkowski2008, WeidlingEtal2009, Ormel2007} already discuss an increase for small mmsizes, the work by \citet{Suyama2008, Okuzumi2012} suggested evolution of
very high porosities to much larger aggregate sizes. This is strongly related to
the properties of the monomers considered. However, the filling factor is the important parameter for 
future collisions as different porosities lead to very different results (sticking, bouncing, fragmentation) as described by \citet{langkowski2008, wurm2005, meisner2012, BlumEtal2006, SchraeplerEtal2012, Meru2013, GeretshauserEtal2011, SchaeferEtal2007}.

In principle, the filling factor can span a large range of between 0 and 1. It seems to be restricted, however, if it is established by collisions. \citet{TeiserEtal2011a}
and earlier work used dust targets prepared manually and in those studies 33\,\% $\pm 1\,\%$ seems to be a limit in filling factor as target parts otherwise chip off. \citet{Meru2013} find in numerical simulations that a filling factor of 37\,\% sets a boundary in numerical simulations of collisions between 1 and 27.5\,m/s in velocity.  
Much higher filling factors can be produced only using omnidirectional compression in the laboratory \citep{GuettlerEtal2009, meisner2012} or in numerical simulations \citep{Seizinger2012}.
The question remains open  if there are typical porosities for 
bodies in protoplanetary disks evolving through collisions, and this is one part of this paper. 

A few experiments at moderate collision velocities exist. In the experiments by
\citet{Kothe2010}, the resulting filling factor reaches up to 40\,\% at 6\,m/s as projectiles
always hit the same spot. A lower value of about 30 to 33\,\% results in \citet{TeiserEtal2011b} and \citet{meisner2012}. A thick layer of dust is grown with random impact sites in those experiments. The experiments reported here extend these two latter works. 

Porosity is important in collisions as it determines the strength of a dust aggregate
and its ability to dissipate energy. In low porosity aggregates particles stick together more rigidly. If the energy of a collision is not sufficient to break contact, the collision will be elastic and will result in bouncing. Bouncing is supposedly the dominant outcome of 
collisions under protoplanetary disk conditions once compact aggregates of mm to cm in size have formed \citep{blum2008}. As bouncing prevents further aggregate growth, \citet{ZsomEtal2010} called this the bouncing barrier. Even if compact aggregates stick to each other the contacts are very weak and the forming aggregates might be destroyed again \citep{JankowskiEtal2012}. Recent experiments on long-term observations of 
a particle system show that the bouncing barrier in a system of mm aggregates is a very robust result \citep{Kelling2013}.

However, if one of the collisional partners is large enough, collision velocities increase and growth is possible again. \citet{wurm2005}, \citet{TeiserWurm2009b}, and \citet{TeiserEtal2011a} carried out collision experiments and showed that a large dusty body in a collision with a submm to mm aggregate gains mass. \citet{TeiserWurm2009b} sketch a model where larger bodies grow at the expense of small particles. \citet{windmark2012} carried out numerical simulations 
and found that growth is possible if a larger seed is introduced into
the system. In another work, \citet{Windmark2012b} as well as \citet{garaud2013} argue that the velocity distribution in a turbulent disk might provide a few seeds by chance if some particles collide
at very low speeds. \citet{JankowskiEtal2012} speculate that aggregates consisting
of larger grains (by chance) might also provide seeds for further growth. 

In this work, we aim to provide quantitative data to supplement numerical models like those in \citet{windmark2012}, as we measured the accretion efficiencies in ``high speed'' dust collisions for the first time.

There is an alternative model to sticking collisions for planetesimal formation based on gravitational instabilities. The basic ideas go back to \citet{GoldreichWard1973} and \citet{Safronov1969}.
A dense dust subdisk forms that eventually collapses due to its own
gravity. \citet{Weidenschilling1989} found that the shear between the dense
midplane and upper layers would create turbulence that would disperse
the particles again and lower the density. However, numerical modeling
of particle motion in turbulent disks in recent years show that turbulence might actually be beneficial to enhance particle densities. Gravitoturbulence and streaming instabilities or density enhancements in pressure maxima or eddies produced otherwise are currently considered one way to jump from cm or dm size to planetesimals without
the hassle of sticking or bouncing collisions \citep{JohansenEtal2007, Klahr2005, dittrich2013, cuzzi2003, YoudinJohansen2007, ChiangYoudin2010}. 

If the needed initial particle size behind this model coincides with the aggregate size at the bouncing barrier, this would be a nice connection between the two mechanisms. If the bouncing barrier is at mm size and instabilities need larger particles, there might still be a need to overcome the bouncing barrier in which case collisional growth would eventually be able to proceed as well \citep{windmark2012}.
In any case, collisions in the dense cluster of particles will also occur in these models. At the typical collision velocities of tens of m/s these collisions will definitly be 
destructive and lead to fragmentation into smaller particles again \citep{SchraeplerEtal2012, Deckers2013}.
The scavenging of fragments by other bodies is eventually of importance in this model as well.

Here, we report on the first experiments with a novel setup that allow a
target to grow by a large number of successive collisions with small dust aggregates. In particular, we quantify an accretion efficiency at high velocity for the first time 
and provide a measure of the porosity evolution of large growing bodies.

\section{\label{sec:Experiments}Experiment basics}
To study collisions of dust grains with velocities up to 71\,m/s, we use a centrifuge, which accelerates submm dust grains in a vacuum chamber. We were able to analyze the small dust agglomerates and their behavior concerning collisions with targets using a high speed camera. Larger and denser dust agglomerates could be produced by the impinging dust particles.  

\subsection{\label{sec:experimental_setup}Experimental setup}
\begin{figure}[h]
\centering
\includegraphics[width=\columnwidth]{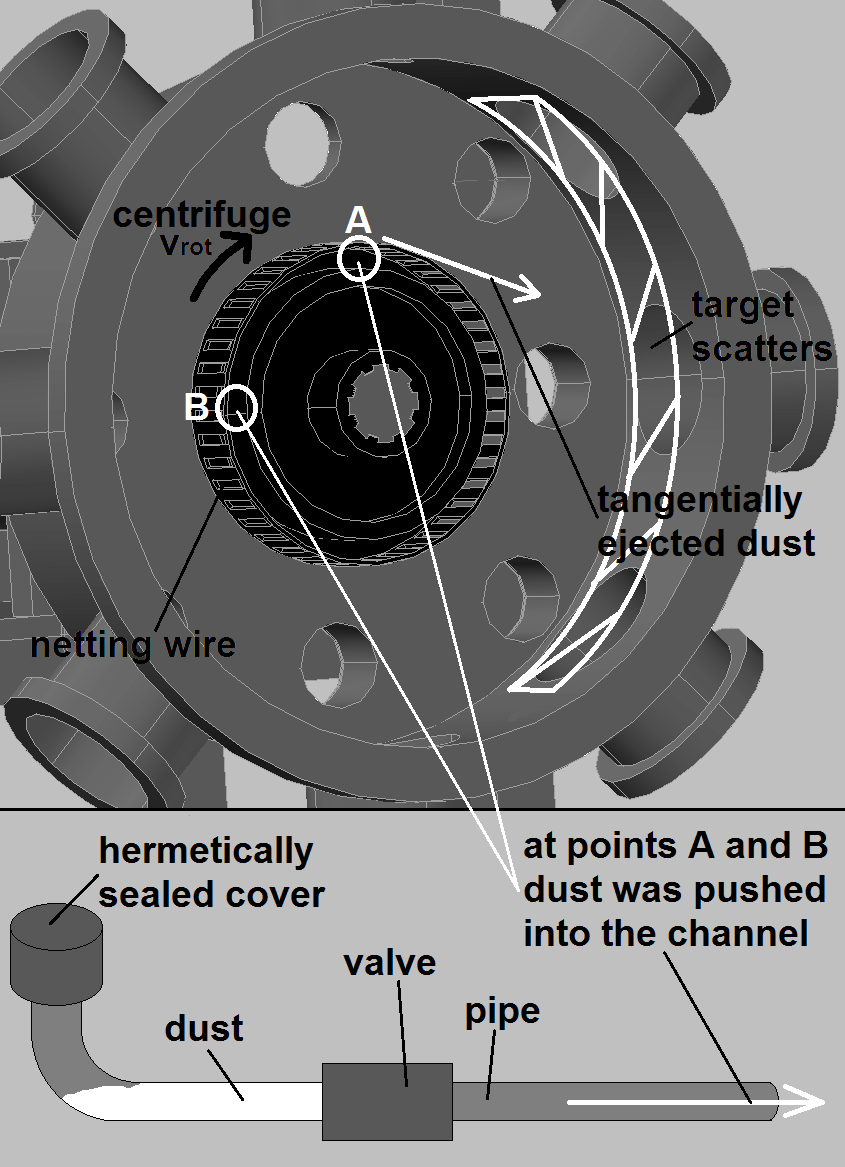}
\caption{A centrifuge rotates inside of a vacuum chamber. Dust is injected at 
points A or B in a sequence of filling a dust reservoir and opening a valve which
sucks in the dust into the centrifuge. Dust aggregates of certain size leave the mesh of the centrifuge tangentially and hit impact areas at the inner surface of the chamber.}
\label{trommel3d}
\end{figure}

The basic element of the setup is the fast rotation of a meshed, hollow cylinder (a centrifuge), placed into a vacuum chamber (Fig. \ref{trommel3d}). Different rotational velocities are realized by using a frequency converter for the engine. 
The centrifuge itself is designed with a narrow, hollow channel at its outer radius.
A fine-meshed netting wire (mesh size: 500\,$\rm\mu$m) is placed at the outside 
of the centrifuge. We injected dust into the centrifuge in a sequence of filling a dust reservoir at the outside of the chamber, which is sucked into the centrifuge if a valve is opened.
The pressure within the chamber is kept between P = 30\,Pa and 80\,Pa. The dust enters the centrifuge and part of it moves toward the mesh where it is tangentially launched toward the inner surface of the vacuum chamber (Fig. \ref{trommel3d}). 
At different positions windows or targets can be placed to observe the  
impacts with a camera at up to 8443\,frames/s. We denote the launch direction as z-axis. The target layer is described by an x-y-plane perpendicular to the z-axis. The y-direction is the line of sight of the camera. Depending on the experiment and details to be studied, we use an observation with continous illumination, flash lamps, and a post-experiment analysis of the target with respect to mass and volume (filling factor). Our dust material is quartz dust with particle sizes ranging from 0.1$\mu$m to 10$\mu$m (80\% are in a range between 1$\mu$m and 5$\mu$m). This quartz dust was used in several previous experiments (\citet{TeiserWurm2009a, TeiserWurm2009b, meisner2012, beitz2011}). We assume a volume filling factor of 0.32 for the launched dust aggregates. This value is typical for locally-compressed dust agglomerates (\citet{TeiserEtal2011b, meisner2012}). However, in this context, the value of the volume filling factor is uncertain and cannot be further quantified. 

\subsection{\label{sec:Vel_engfreq}Velocity calibration}

During the experiments an accumulation of dust could be traced on the inner surface of the vacuum chamber at a maximum of 135\,$^{\circ}$ away from the injection point along the direction of rotation. As impacts and impacting particles have different velocity regimes, 
we determined the impact velocity for a free-flying dust particle first. This allows us to set the collision velocity and is correlated to the frequency of the engine converter.. 
As we cannot trace every incoming dust aggregate, we consider these calibrated values to be impact velocities. 

\begin{figure}[h]
\centering
\includegraphics[width=\columnwidth]{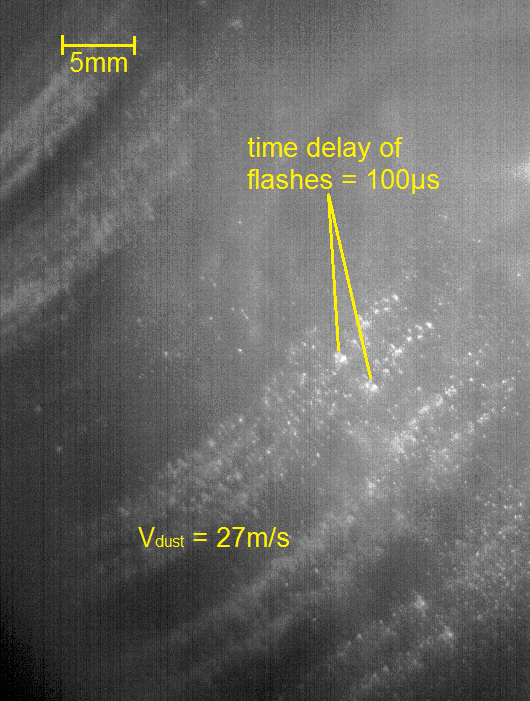}
\caption{Double exposure image of dust particles launched by the centrifuge. 
Measuring the lengths between two similar features at a known delay time gives
the velocity of the particles. The direction to the centrifuge is top-left.}
\label{Hoellevel4}
\end{figure}

\begin{figure}[h]
\centering
\includegraphics[width=\columnwidth]{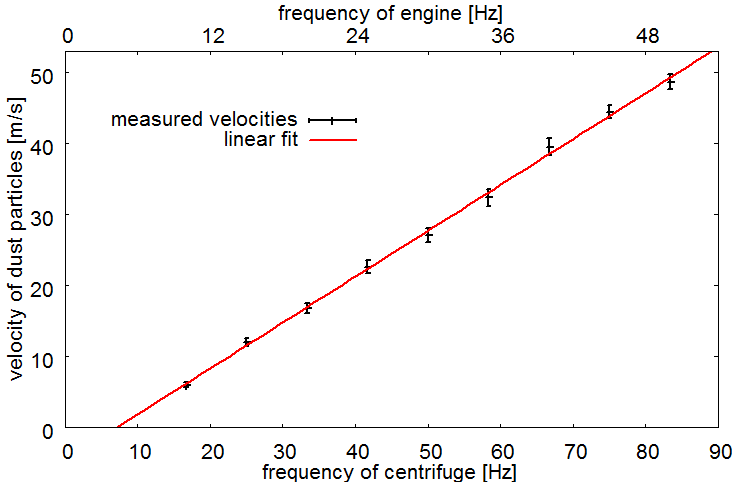}
\caption{Particle velocities follow a linear dependence on engine/centrifuge frequency.}
\label{freqvel}
\end{figure}

We used two flash lamps for a double exposure of dust volleys from the injection. The time delay between the two pulses varied between 60\,$\rm\mu$s - 200\,$\rm\mu$s.      
Fig. \ref{Hoellevel4} shows an example of this. 
By measuring the lengths between the same features and given a predetermined time delay of the flash lamps we get the (collision) velocities of the dust aggregates for different engine frequencies. In Fig. \ref{freqvel} our measured velocities are plotted against the rotation frequency of the centrifuge as well as the converter frequency of the engine. 
The dust is expelled from the centrifuge only for frequencies larger than 8\,Hz and the
linear fit is offset, but not crossing the origin. Otherwise, as expected, there is a linear increase in measured velocities $v$ in [m/s] of the launched dust with frequency $f$ of the centrifuge given as: 
\begin{equation}
v(f) = 0.65\,[\rm{m}] \cdot f - 4.64\,[\rm{m/s}].
\label{velequation}
\end{equation} 
This dependency is used to calculate a collision velocity for 
a preset converter frequency, which is accurate to about 1 m/s.

\subsection{\label{sec:Size_dust}Aggregate mass}
\begin{figure}[h]
\centering
\includegraphics[width=\columnwidth]{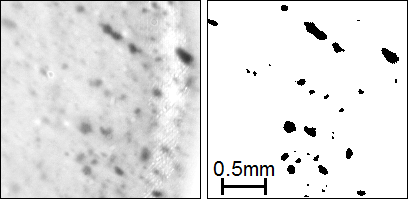}
\caption{Example of a bright field image of dust launched from the centrifuge (left). 
After processing we use the binary image (right) for further analysis.}
\label{hoellemik}
\end{figure}

\begin{figure}[h]
\centering
\includegraphics[width=\columnwidth]{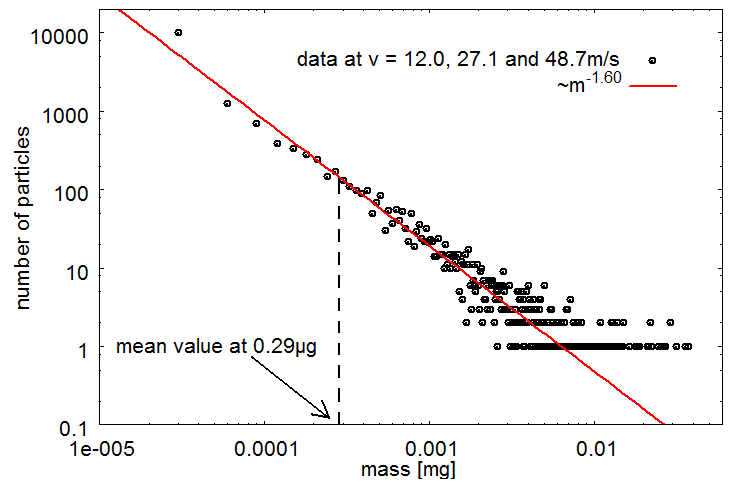}
\caption{Mass distribution of dust launched by the centrifuge through meshes with a size of 500\,$\mu$m at the ejection velocities of 12.0\,m/s, 27.1\,m/s, and 48.7\,m/s. Because the mass distribution does not depend on the ejection velocity all data are combined into one total distribution. The mass distribution decreases to the power of 1.60 and the mean value of the launched masses is located at 0.29\,$\mu$g.}
\label{loghoellemik}
\end{figure}

\begin{figure}[h]
\centering
\includegraphics[width=\columnwidth]{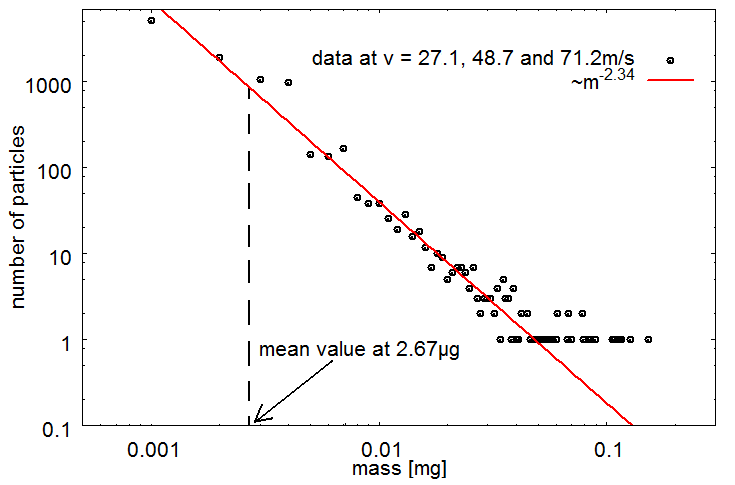}
\caption{Mass distribution of dust launched by the centrifuge at the ejection velocities of 27.1\,m/s, 48.7\,m/s and 71.2\,m/s. The mass distribution decreases to the power of 2.34 and the mean value of the launched masses is located at 2.67\,$\mu$g.}
\label{loghoellemik267}
\end{figure}

To determine the particle size and mass of particles produced by the centrifuge, we imaged particles in higher resolution. For the analysis at engine frequencies of 15\,Hz, 30\,Hz, and 50\,Hz (12.0\,m/s, 27.1\,m/s, and 48.7\,m/s respectively), we counted between 2400 and 7800 particles.  
A typical image before and after processing (binary image) is seen in Fig. \ref{hoellemik}. The binary images only contain information about particles that are located in the focal plane. We set the black/white level of the binary images manually.     
Based on the two-dimensional images, we calculated the radius of an equivalent cross-section sphere for a particle. We multiplied the volume with a density of 2.6\,g/cm$^{3}$ and a volume filling factor, so as to get the particle mass. We assumed a volume filling factor of $0.32$ for the dust particle which is typical for locally compressed dust agglomerates \citep{meisner2012, TeiserEtal2011b}.
For small particles, we reach the resolution limit of the optical system. At very large sizes only few aggregates exist. In between these sizes, the data follow a power law. 
All distributions are roughly proportional to $m^{-3/2}$.
A systematic change of the distributions with velocity cannot be seen.
Hence, the mass distributions at velocities of 12.0\,m/s, 27.1\,m/s and 48.7\,m/s are added to one total distribution in Fig. \ref{loghoellemik}.
The total distribution is proportional to $m^{-1.6}$.  
The average value for the particle mass is 0.29\,$\mu$g. This corresponds to an average particle radius of 43.4\,$\mu$m. 
We also carried out a similar set of experiments, but with larger mesh size of the centrifuge. This produced larger impacting aggregates with an average mass of 2.67\,$\mu$g (average particle radius of 91.5\,$\mu$m), approximately ten times more massive. A systemic change of the distributions with velocity cannot be seen. In Fig. \ref{loghoellemik267}, the mass distributions at velocities of 27.1\,m/s, 48.7\,m/s, and 71.2\,m/s are added to one total distribution. It is proportional to $m^{-2.34}$.  

\section{Impact experiments}

With the means to generate a beam of small aggregates at different velocities with a typical mass of 0.3 \,mg at a size of about 45 \,$\mu$m, we carried out impact experiments. We placed targets at different positions at the inner wall of the vacuum chamber where they are continuously hit by dust aggregates.

\subsection{\label{sec:coef_rest}Coefficient of restitution}
\begin{figure}[h]
\centering
\includegraphics[width=\columnwidth]{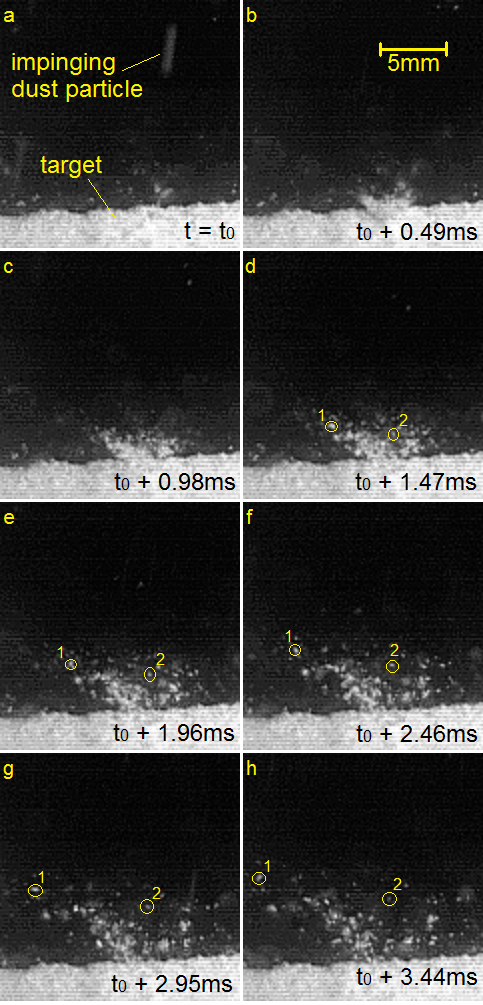}
\caption{Aggregate impact at 27.1\,m/s. The frame rate was set to 2036\,frames/s. The impacting projectile is a stretched trail caused by an exposure time of 55\,$\rm{\mu}$s. As an example of ejecta the positions of two fragments (1 and 2) are marked on 
subsequent images.}
\label{coefrest5}
\end{figure}

A round plastic disk was placed as a target at the bottom of the target zone in Fig. \ref{trommel3d}. Dust accumulates on the target due to direct sticking and reaccretion of ejecta by gravity. The influence of reaccretion is studied on targets further up in the target zone where gravity does not return ejecta. This is described later in this article. Here, we study the ejection process of fragments after an impact 
onto an accumulated dust bed. We recorded the trajectories of single fragments with a high-speed camera at 2036\,frames/s. A typical impact scenario with marked fragments is shown in Fig. \ref{coefrest5}.     

For bouncing collisions, the energy loss can be quantified by the ratio between the velocity after and before bouncing. Similarly, the ejecta can also be characterized by a coefficient of restitution, which we define here as
\begin{equation}
C_{R} = \frac{v_{f}}{v_{c}}
\end{equation}    
with collision velocity $v_{c}$ and ejecta velocity $v_f$. As can be seen in Fig. \ref{coefrest5}, the dust target is placed almost perpendicular (up to 7\,$^{\circ}$) to the impinging dust particles on the target. To a good approximation the x-component of the collision velocity can be neglected.   
The fragment velocity $v_{f}$ can be calculated as
\begin{equation}
v_{f} = \sqrt{v_{x}^{2} + v_{y}^{2} + v_{z}^{2}} = \sqrt{2 \cdot v_{x}^{2} + v_{z}^{2}}   
\end{equation}    
In this case, the z-direction coincides with the vertical direction and $v_x$ and $v_y$ are the horizontal velocities and $v_z$ is the vertical ejecta velocity with respect to the target surface. Since we only use one camera, $v_{y}$ is unknown. We assume that it is independent of, but on the same order as, $v_{x}$.  

The horizontal velocity $v_x$ is easy to deduce from the images as it does not
change with time. In vertical direction, gravity has to be considered acting with acceleration g = 9.81\,m/s$^{2}$ and the velocity is given as
\begin{equation}
v_{z} = \frac{\Delta z - \frac{1}{2}g\Delta t^{2}}{\Delta t}.
\end{equation}  

\begin{figure}[h]
\centering
\includegraphics[width=\columnwidth]{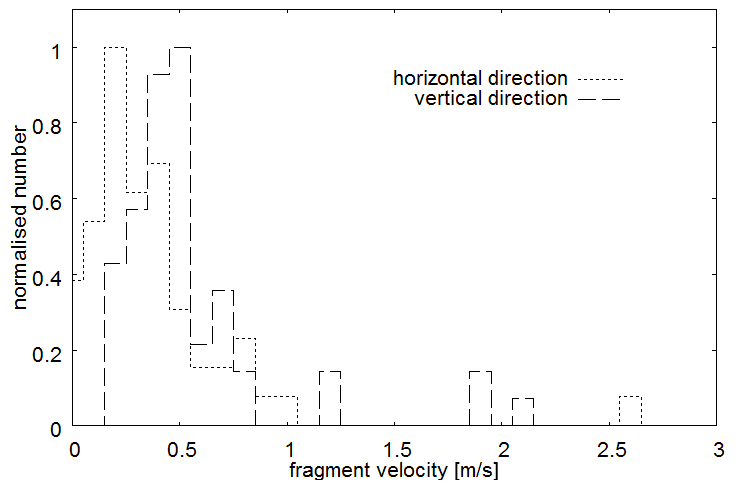}
\caption{Ejecta velocities for collision velocity $v_{c}$ = 27.9 $\pm$ 0.9\,m/s  }
\label{fragment30}
\end{figure}

\begin{figure}[h]
\centering
\includegraphics[width=\columnwidth]{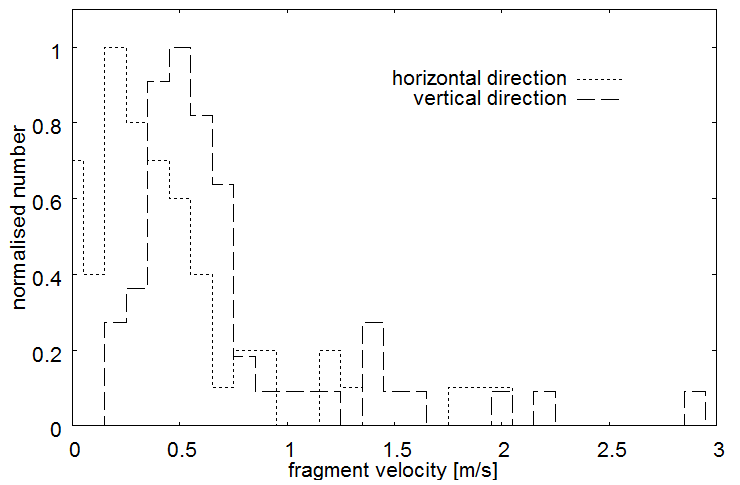}
\caption{Ejecta velocities for collision velocity $v_{c}$ = 49.5 $\pm$ 1.0\,m/s.  }
\label{fragment50}
\end{figure}

We determined fragment velocities for 27.1\,m/s and 48.7\,m/s collision velocities. In total, we analyzed 56 and 58 fragment velocities, respectively, in this work and they are illustrated in figures \ref{fragment30} and \ref{fragment50}.

At 27.9\,m/s the average horizontal velocity and the standard deviation are v$_{x}$ = 0.325 $\pm$ 0.241\,m/s and vertical ejecta velocities are v$_{z}$ = 0.554 $\pm$ 0.394\,m/s. At 49.5\,m/s the average values are v$_{x}$ = 0.465 $\pm$ 0.453\,m/s  and v$_{z}$ = 0.721 $\pm$ 0.510\,m/s. 

\begin{figure}[h]
\centering
\includegraphics[width=\columnwidth]{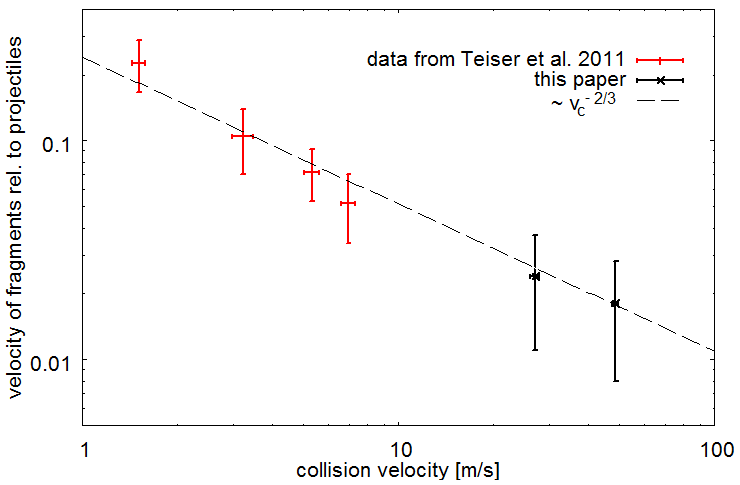}
\caption{Coefficients of restituion for ejecta of a $\rm SiO_2$ aggregate impacts. The dashed line is a power law of $C(v_{c}) = 0.24 \cdot v_{c}^{-2/3}$}
\label{coefrestall}
\end{figure}

We used these values to calculate a coefficient of restitution.
We compare the values to measurements by \citet{TeiserEtal2011b} at lower collision velocities on particles of the same composition. Those aggregates were somewhat larger (250\,$\mu$m in diameter). This is shown in Fig. \ref{coefrestall}.
The data can be well approximated by a power law within the range of data
\begin{equation}
C(v_c) = 0.24 \cdot {v_c}^{-2/3}.
\end{equation} 
We do not attempt to motivate this power law here but only take it as one of the most simple analytic expressions that fits the data between 1\,m/s and 50\,m/s. As long as we use submmsize dust projectiles, the fragment velocity in relation to the velocity of the projectiles fits this analytic expression very well, independent from variations in particle sizes.    
 
There have to be deviations to very small and very large velocities. At low velocity, the coefficient of restitution cannot be larger than 1. At high velocity is is likely that $C$ levels off once the aggregate is completely fragmenting to individual grains and no more energy can be dissipated by breaking contacts.

\subsection{\label{sec:mass_highspeed}Growing targets at high speed}
\begin{figure}[h]
\centering
\includegraphics[width=\columnwidth]{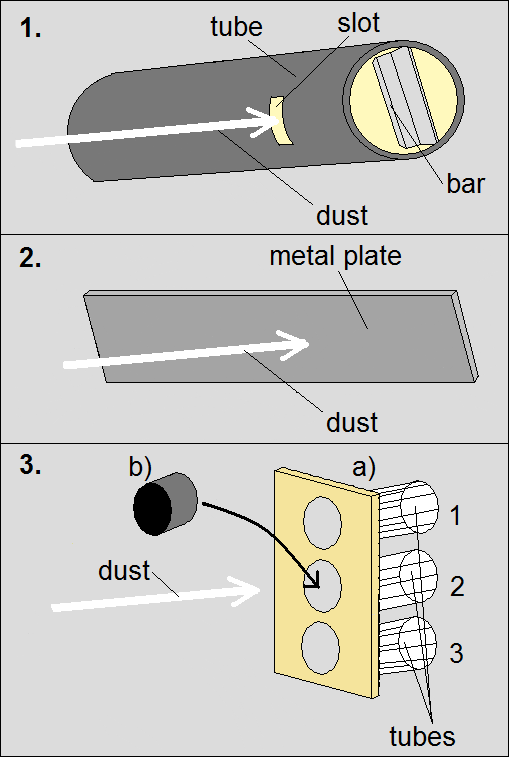}
\caption{1. Target configuration: dust projectiles enter a small aperture within a tube and impact onto a thin bar. 2. Target configuration: dust projectiles impact an inclined metal plate above the centrifuge directly (no reaccretion due to gravity) 3. Target configuration: in a first step (a), the local mass distribution is probed by measuring the masses inside neighboring tubes, which collect all of the dust. In a second step (b), the mass of an aggregate grown on an exposed target is measured with calibration tubes collecting dust at the same time.}
\label{acceff_new}
\end{figure}

As illustrated in Fig. \ref{acceff_new}, dust projectiles impact targets in three different settings.  

\begin{figure}[h]
\centering
\includegraphics[width=\columnwidth]{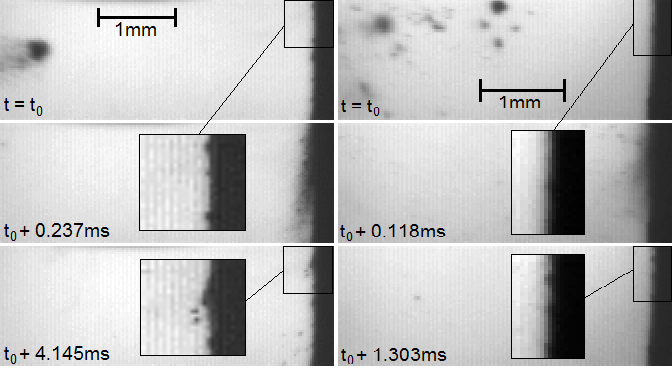}
\caption{On the left is an example of simoultaneous mass accretion and erosion. As can be compared in the highlighted frames before and after an impact of a fast dust particle (v = 48.7\,m/s) 225\,$\mu$m in diameter, the small hill on top is lost. At the same time, a new small hill is built up again where the particle impacted the target. On the right is an example for mass accretion. After a multiple impact of dust particles $<$ 110\,$\mu$m in diameter (v = 48.7\,m/s), new small hills are built up.}
\label{massacc1}
\end{figure}

\textbf{1.  Small target:}
In a first setup of a target the goal was to detect growth or erosion of a target in individual collisions. To prevent particles from returning by gravity, the target is placed higher than the centrifuge at an incline. Injection of dust is placed in position B in Fig. \ref{trommel3d}. Every nonsticking particle does not return to the target. As our first target, we chose a 2\,mm metal bar, which was mounted inside a tube (Fig. \ref{acceff_new}.1).
Dust projectiles entered the tube only through an aperture slightly larger than
the target. 
Otherwise, the tube shielded the target from opaque and unfocused ``dust storms'', which makes it impossible to observe single collisions on the edge of the bar.  We set the frame rate to 8443\,frames/s. 
In the bright-field image (Fig. \ref{massacc1}), we could identify the direct addition of mass at the bar`s edge as well as removal of dust. 
After the experiment, a small steep dust crest was created on the target. The importance of these experiments is as follows. Without exact number densities, which are hard to observe, and without at least simple modeling, we cannot rule out {\it a priori} that ejected particles are not hit by later projectiles while
airborne. Such secondary collisions close to the target might lead to follow-up collisions at different speeds, which would spoil the analysis of the experiment. We see only very few such secondary collisions and they cannot be related to target growth. Therefore, this part of the impact experiments show that
the densities in the experiments are sufficiently low that impacts can be regarded as individual events. 

\begin{figure}[h]
\centering
\includegraphics[width=\columnwidth]{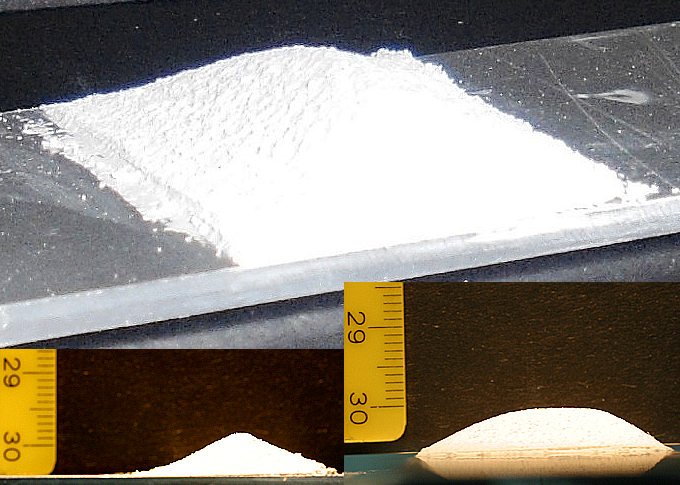}
\caption{One of the produced dust agglomerates built by collisions of dust grains on a metal plate with $v_{c}$ = 48.7\,m/s. Its length is about 3\,cm and the height is 0.6\,cm. }
\label{agglom}
\end{figure}

\textbf{2. Large targets}
With the knowledge that there is insignificant interaction between the ejecta cloud and the projectile cloud, larger targets can be grown to analyze the volume filling factor
of the forming aggregates. 
Instead of the tube with the thin bar, we now only place a free metal plate into the path of the projectiles (Fig. \ref{acceff_new}.2). As in the setup before the target was placed higher than the centrifuge to prevent reaccretion. We
used a velocity of about 50 m/s here and produced two grown dust aggregates.
The plate with the grown dust was carefully removed from the vacuum chamber
after growing for some time. 
The dust aggregates had sizes of about 3\,cm in length and 0.6\,cm in height as can be seen in Fig. \ref{agglom}. 

To determine the volume filling factor, mass and volume have to be known. The mass is easily measurable to high accuracy. The volume was determined with a new procedure. By illuminating the agglomerate from above and shadowing a part of the target we could image the projected cross-sections along the terminator. Within the illuminated cross-sections, pixels were counted and were calculated into an area. The terminator was moved incrementally in 1\,mm steps. In relation to 3\,cm length of the agglomerates, we could calculate volumes for equidistant slices that are thin enough that the volume of the whole agglomerate could be approximated very well. This method is comparable in precision to the measurement of volumes in \citet{TeiserEtal2011b}. We chose the new method because the method of \citet{TeiserEtal2011b} has its shortcomings for asymmetric targets. This way volume filling factors of 0.38 and 0.39 were found. The uncertainties due to the volume determination are 5\%.      

If we place the target at the bottom of the chamber, reaccretion can lead to slow secondary collisions. These rebounding ejecta might build a high porosity top layer on a target. Subsequent high speed impacts by the next set of projectiles might compact this layer, but the total volume filling factor
might still be smaller as energy is dissipated in this process. This was not observed in experiments at low speed by \citet{TeiserWurm2009a}, but might be present at high speed collisions. 

\begin{figure}[h]
\centering
\includegraphics[width=\columnwidth]{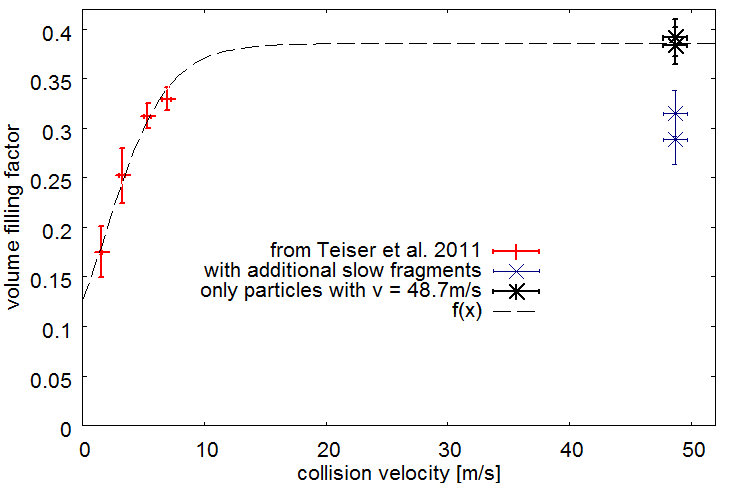}
\caption{Volume filling factors of aggregates. See text for details on f(x) equation (\ref{equ:vff}).}
\label{vffges2}
\end{figure}

To estimate such effects we grew dust aggregates on targets that were placed below the rotating centrifuge. In this setup configuration, ejecta from fast collisions could partly settle down onto the developing dust aggregate. The volume filling factors measured were 0.29 and 0.32. Fig. \ref{vffges2} shows the results with earlier findings at lower velocities by \citep{TeiserEtal2011b}.     
To give an analytic expression we fit the following function to our data:
\begin{equation}
f(x) = \frac{a}{b + e^{-c \cdot x}}
\label{equ:vff}   
\end{equation}            
with a = 0.19, b = 0.49, c = 0.39\,s/m. 
At 48.7\,m/s for the impact velocity of projectiles, we measured two values of filling factors for both positions of the target (with and without gravitational reaccretion). We found slightly higher filling factors at collision velocities of nearly 50\,m/s because for that case the position of the target (above the centrifuge and hence, the surface adjusted downwards) did not allow the reaccretion of ejecta. The dust agglomerate grew almost entirely by direct sticking of dust projectiles at this high velocity. That is in contrast to \citet{TeiserEtal2011b}, where the agglomerates grew with a mixture of both effects: direct sticking and reaccretion. This case is simulated when changing the position of the target (lower side of the chamber). The saturation level of the filling factor assumed around 0.32 by \citet{TeiserEtal2011b} is within the error bars of the two datapoints with reaccreted ejectas.     

\textbf{3. Accretion efficiency:}
It is not only important to know {\it that} large bodies grow in collisions with smaller particles, it is also necessary to know the accretion efficiency, which is not well constrained so far.
As accretion efficiency we define the ratio $\epsilon = m_{stick} / m_{total}$ with $m_{stick}$ as mass sticking to the target (mass of the dust pile) after a certain time and $m_{total}$ being the total mass that impacted during this time. To measure the total impacting dust, we placed three tubes in a row next to each other. (see Step a, Panel 3 of Fig. \ref{acceff_new}).  
Dust entering the tube stays in the tube so a total mass of dust hitting the 
tube opening can be measured. In a number of experimental runs we measured the ratio of masses between the different tubes to calibrate the difference of impacting mass with varying tube locations. This way measuring the mass in one tube yields a measure of the total impacting mass in the other tubes or $m_{total}$ by adding a calibration factor. This factor was determined to an accuracy of 8.7\%, 3.5\%, 3.1\%, and 7.2\% for velocities of 27.9\,m/s, 38.7\,m/s, 49.5\,m/s, and 71.2\,m/s, respectively. In a second set of runs, the center tube was replaced by a small target with the same diameter as the original tube opening (see Fig. \ref{acceff_new}, panel 3, step b). An aggregate grows on the target and its mass $m_{stick}$ is measured after a certain time. For a given collision velocity (calculated with equation (\ref{velequation}) from the velocity calibration) between six and ten calibration runs and six to seven impact runs were carried out. The accreted masses, the single accretion efficiencies and measurement errors $\sigma$ due to mass determination and calibration are given in Table \ref{measuredmass}. 
\begin{table}[h]
\centering
\caption{The mass which sticks on a target ($m_{stick}$) and the mass, which is launched toward the target ($m_{total}$) at a given impact velocity. The ratio of $m_{stick}$ and $m_{total}$ gives the accretion efficiency $\epsilon$. The parameter $\sigma$ is the error due to mass determination and calibration. The average value of impacting masses was 0.29\,$\mu$g.}
\begin{tabular}{|c|c|c|c|c|}
\hline
v [m/s] &  $m_{stick}$ [mg] & $m_{total}$ [mg] & $\epsilon$ [\%] &  $\sigma$ [\%] \\ \hline\hline
27.9           &         10       &      20           &			50.0			 &  8.1     \\
							 &				 6        &      9            &			66.7			 &  13.9    \\       
							 &				 5        &      8          	&			62.5			 &  15.6  	\\
							 &				 11       &      16          	&			68.8			 &  10.8  	\\
							 &				 7        &      14          	&			50.0			 &  9.8   	\\
							 &				 4        &      7           	&		  57.1			 &  16.2  	\\ \hline
38.7           &         18       &      56          	&			32.1       &  2.4   	\\
							 &				 8        &      28           &			28.6       &  4.0   	\\       
							 &				 10       &      25           &			40.0       &  4.7   	\\
							 &				 18       &      37          	&			48.6       &  3.8   	\\
							 &				 12       &      33           &			36.4       &  3.6   	\\
							 &				 15       &      37          	&			40.5       &  3.7   	\\ \hline
49.5           &         43       &      133         	&			32.3       &  1.4   	\\
							 &				 39       &      101          &		  38.6       &  1.7   	\\       
							 &				 32       &      104          &			30.8       &  1.5   	\\
							 &				 64       &      181          &			35.4       &  1.4   	\\
							 &				 44       &      103          &			42.7       &  1.8  		\\
							 &				 64       &      164          &			39.0       &  1.5   	\\ \hline 
71.2           &         52       &      547         	&			9.5        &  0.7   	\\
							 &				 67       &      229          &		  29.3       &  2.0   	\\       
							 &				 61       &      258          &			23.6       &  1.5   	\\
							 &				 59       &      318          &			18.6       &  1.3   	\\
							 &				 41       &      235          &			17.4       &  1.2  		\\
							 &				 7        &      51           &			13.7       &  2.2   	\\ \hline		 			 
\end{tabular}
\label{measuredmass}
\end{table}  
For intermediate and high velocities the scatter in individual experiment runs is larger than the individual measurement errors. The average accretion efficiencies and the uncertainties based on the standard deviation are given in Table \ref{acceff}. These values can be regarded as self-consistent accretion efficiencies for an evolving target if the dust layer is thick enough that initial effects of impacting the
plastic substrate can be neglected. The thickness of the dust layer can be calculated from the accreted mass $m$ as 
\begin{equation}
d = \frac{m}{\pi s^{2} \rho \Phi}    
\end{equation}  
where $s$ = 5\,mm is the radius of the target plate,
$\rho =$ 2.6\,g/cm$^{3}$ is the dust density, and the assumed filling factor $\Phi$ is 0.38. A 10 mg mass gain corresponds to a thickness of 0.13\,mm. This is much larger than the particle size and is also several times the projectile size. We consider this as large and argue that we measured the self-consistent growth of a dust target for masses larger than 10 mg. Most measured masses were even larger. One exception are the values at 27.9\,m/s. Due to technical limitations only a small amount of dust mass could be measured here and the dust layer is on the order of the projectile size. Instead, we measure the sticking fraction of dust on the dustless plastic target here and find no self-consistent growth that might include erosion from the dust layer. We have yet to find a better way to quantify this systemic difference, therefore this accretion efficiency might be systemically too high. For 48.7 m/s the dust flux was largest and the accretion efficiencies were large. This allowed the buildup of targets extended enough to determine their filling factors by direct mass and volume measurements.        
\begin{table}[h]
\centering
\caption{Accretion efficiencies of dust (mean mass: 0.29\,$\mu$g) at given impact velocity.}
\begin{tabular}{|c|c|c|}
\hline
velocity [m/s] & accretion efficiency [\%] & 1 $\sigma$ [\%] \\ \hline\hline
27.9                    &          58.4    &     9.1         \\
38.7                    &          38.7    &     7.0         \\
49.5                    &          36.8    &     4.7         \\
71.2                    &          18.3    &     6.8         \\ \hline
\end{tabular}
\label{acceff}
\end{table}  

In a second series of experiments we measured the accretion efficiencies of impacting dust particles, which are a magnitude larger in mass (mean mass = 2.67\,$\mu$g) with the same procedure as explained above. We produced larger particles by removing the fine-meshed netting wire (mentioned in Sect. \ref{sec:experimental_setup} and in Fig. \ref{trommel3d}) at the outside of the centrifuge. Again, the determined masses, the single accretion efficiencies and measurement errors are given in Table \ref{measuredmass2}. One additional difference in view of the low masses measured at low velocity for small aggregates was that the plastic target was replaced by a precompacted dust target of 0.48 in filling factor. Therefore, the first impacts already took place on dust targets. However, the dust flux and accreted masses were large enough here to consider the growth as self consistent.  
\begin{table}[h]
\centering
\caption{Same as Table \ref{measuredmass}, but for larger projectiles (2.67\,$\mu$g).}
\begin{tabular}{|c|c|c|c|c|}
\hline
v [m/s] &  $m_{stick}$ [mg] & $m_{total}$ [mg] & $\epsilon$ [\%] &  $\sigma$ [\%] \\ \hline\hline
27.9           &         31       &      166          &			18.7			 &  7.2     \\
							 &				 59       &      379          &			15.6			 &  7.1     \\       
							 &				 44       &      140          &			31.4			 &  7.2  	  \\
							 &				 52       &      184          &			28.3			 &  7.1  		\\
							 &				 47       &      162          &			29.0			 &  7.2   	\\
							 &				 43       &      176          &		  24.4			 &  7.1  		\\ \hline
38.7           &         46       &      222          &			20.7       &  8.7   	\\
							 &				 31       &      170          &			18.2       &  8.7   	\\       
							 &				 21       &      372          &			5.6        &  8.7   	\\
							 &				 43       &      194          &			22.2       &  8.7  		\\
							 &				 41       &      209          &			19.6       &  8.7  		\\
							 &				 45       &      251          &			17.9       &  8.7  		\\ \hline
49.5           &         38       &      342         	&			11.1       &  6.9  		\\
							 &				 46       &      203          &		  22.7       &  7.0  		\\       
							 &				 27       &      165          &		  16.4       &  7.0  		\\
							 &				 37       &      234          &			15.8       &  6.9  		\\
							 &				 23       &      212          &			10.8       &  7.0   	\\
							 &				 46       &      246          &			18.7       &  6.9  		\\ \hline 
71.2           &         36       &      357         	&			10.1       &  5.2  		\\
							 &				 14       &      494          &		  2.8        &  5.2   	\\       
							 &				 49       &      689          &	  	7.1        &  5.2   	\\
							 &				 48       &      687          &			7.0        &  5.2   	\\
							 &				 45       &      669          &			6.7        &  5.2    	\\
							 &				 30       &      532          &			5.6        &  5.2  		\\ \hline 	 							 	\end{tabular}
\label{measuredmass2}
\end{table}  
As calculated for the smaller impacting particles, the average accretion efficiencies and the scatter in terms of a 1 sigma deviation are also given for the larger particles in Table \ref{acceff2}.				 
\begin{table}[h]
\centering
\caption{Accretion efficiencies of dust (mean mass: 2.67\,$\mu$g) at given impact velocity.}
\begin{tabular}{|c|c|c|}
\hline
velocity [m/s] & accretion efficiency [\%] & 1 $\sigma$ [\%] \\ \hline\hline
27.9           &          24.6              &     6.9        \\
38.7           &          17.4              &     6.9        \\
49.5           &          15.9              &     5.3        \\
71.2           &          6.6               &     3.2        \\ \hline
\end{tabular}
\label{acceff2}
\end{table}  
							 
\begin{figure}[h]
\centering
\includegraphics[width=\columnwidth]{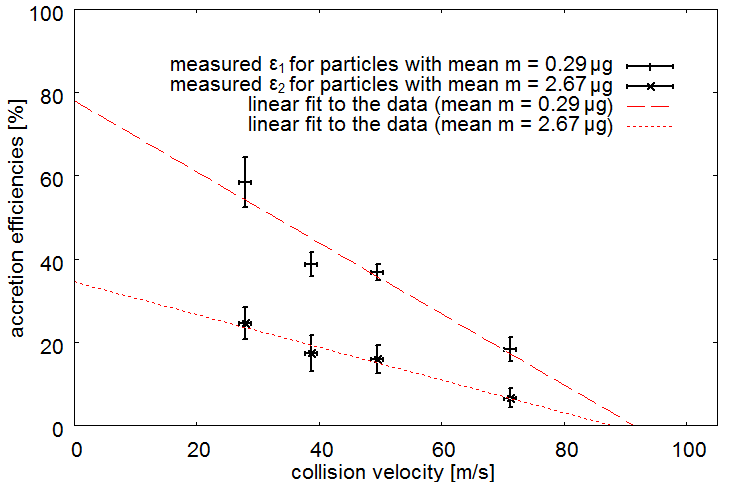}
\caption{Accretion efficiencies $\epsilon_{1}$ and $\epsilon_{2}$ of growing dust agglomerates by impinging dust particles (mean masses m: 0.29\,$\mu$g and 2.67\,$\mu$g) in dependency of their collision velocities.}
\label{acceff1}
\end{figure}

\begin{figure}[h]
\centering
\includegraphics[width=\columnwidth]{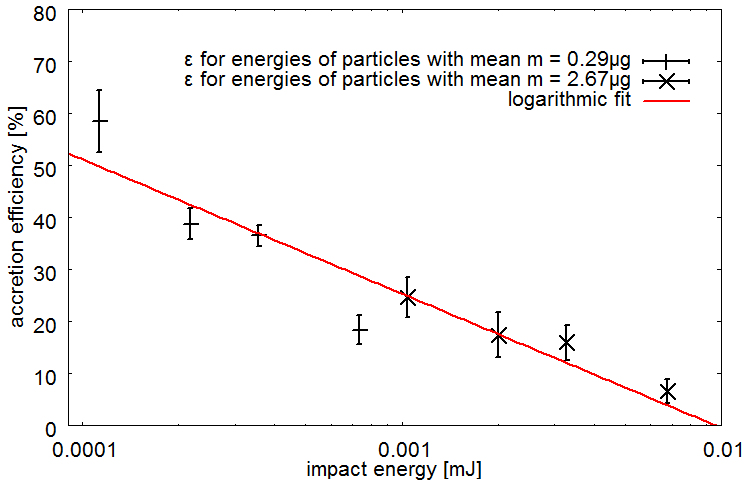}
\caption{Accretion efficiencies over impact energy of particles with different values of their mean mass.}
\label{acceffen}
\end{figure}

The determined values for the accretion efficiency show a distinct tendency to be reduced with increasing collision velocity. We fit our data from Tables \ref{acceff} and \ref{acceff2} with linear functions (Fig. \ref{acceff1}) each, as the most simple function in agreement with the data. The accretion efficiencies are hereby given as:
\begin{equation}
\epsilon_{1}(v) = -0.85 \cdot v + 78    
\end{equation}  
for the colliding particles with a mean mass of 0.29\,$\mu$g and
\begin{equation}
\epsilon_{2}(v) = -0.39 \cdot v + 35    
\end{equation}
for the colliding particles with a mean mass of 2.67\,$\mu$g   
with $\epsilon_{1}$ and $\epsilon_{2}$ as accretion efficiencies in [\%] and the velocity $v$ in [m/s]. 
If collision velocities are increasing beyond 90\,m/s, no dust aggregates should stick onto the target anymore. Growing dust agglomerates are likely being destroyed by impinging dust, which possesses too much kinetic energy. 
It might be worth considering the dependence of the accretion efficiency on the impact energy. Therefore, we used the mean values of mass of the dust particles to calculate their kinetic energies at both measurement runs; their different accretion efficiencies are plotted in Fig. \ref{acceffen}. The determined values for the accretion efficiency show a distinct tendency to be reduced with increasing collision energies.
The decrease of the accretion efficiency with increasing collision energy for both datasets can well be described by one logarithmic dependence as:
\begin{equation}
\epsilon(E_{c}) = -11.2 \cdot ln(E_{c}) - 52    
\end{equation}  
with $\epsilon(E_{c})$ as accretion efficiency in [\%] and the collision energy $E_{c}$ in [mJ].

\section{\label{sec:conclusions}Conclusions}

The basic experimental conclusions of this work are straightforward.
If dust aggregates of about $50\,\mu m$ diameter, consisting of $\rm SiO_2$ particles of a few micrometer in size, collide continuously with a larger body at random locations on its surface, we find the following:

\begin{itemize}
\item The larger target body growth at the expense of the smaller projectile aggregate.

\item The accretion efficiency for impacting dust particles with a mean mass of 0.29\,$\mu$g decreases from 58\,\% to 18\,\% between 27\,m/s and 71\,m/s.

\item The accretion efficiency for impacting dust particles with a mean mass of 2.67\,$\mu$g decreases from 25\,\% to 7\,\% between 27\,m/s and 71\,m/s.

\item The accretion efficiency for both data sets can be described by a logarithmic decrease with impact energy from 58\,\% to 7\,\%  at impact energies from 0.1\,$\mu$J and 10\,$\mu$J.

\item The volume filling factor of the target will evolve to 0.38 at 49\,m/s.

\item The volume filling factor increases from 0.32 to 0.38 between 7\,m/s and 49\,m/s.

\item The volume filling factor at high speed is sensitive to the same 
order of a few \% to the slow reaccretion of dust (by gravity here) and subsequent compaction 
in following collisions with a volume filling factor being 0.30.
\end{itemize}

From earlier experiments at high speed by \citet{wurm2005} of mm aggregates impacting up to 25\,m/s, it was known that accretion efficiencies larger than 30\% occurs. This is still true for aggregates of more than an order of magnitude smaller. \citet{TeiserWurm2009b} showed that submm particles are likely to stick up to 60\,m/s, but these results were based on individual collisions. The experiments here show that this is indeed the case and we give quantitative values for volume filling factors of a self-consistently produced target. 

Our results support the idea of particle growth of larger bodies through collisions as suggested by \citet{TeiserWurm2009b} and shown in a model by \citet{windmark2012} to work if a bouncing barrier is present. Also, in a dense environment of clumps of cm or dm aggregates forming in an instability scenario \citep{ChiangYoudin2010}, these findings are applicable as they provide a means to recollect the debris of a destructive collision at high speed. Either way large bodies are very efficient scavengers in protoplanetary disks which, among other things, is certainly a dominating process in early phases of planet formation.

\begin{acknowledgements}
      This work is funded by the 
      \emph{Deut\-sche For\-schungs\-ge\-mein\-schaft, DFG\/ as part of the research group FOR 759}. 
      We thank Manfred Aderholz, Ulrich Visser, and the mechanical workshop for the manufacture of the centrifuge.
\end{acknowledgements}

\bibliographystyle{aa} 
\bibliography{holey} 

\begin{thebibliography}{65}
\expandafter\ifx\csname natexlab\endcsname\relax\def\natexlab#1{#1}\fi

\bibitem[{{Aumatell} \& {Wurm}(2011)}]{Aumatell2011}
{Aumatell}, G. \& {Wurm}, G. 2011, \mnras, 418, L1

\bibitem[{{Beitz} {et~al.}(2011){Beitz}, {G{\"u}ttler}, {Blum}, {Meisner},
  {Teiser}, \& {Wurm}}]{beitz2011}
{Beitz}, E., {G{\"u}ttler}, C., {Blum}, J., {et~al.} 2011, \apj, 736, 34

\bibitem[{{Blum} {et~al.}(2006){Blum}, {Schr{\"a}pler}, {Davidsson}, \&
  {Trigo-Rodr{\'\i}guez}}]{BlumEtal2006}
{Blum}, J., {Schr{\"a}pler}, R., {Davidsson}, B., \& {Trigo-Rodr{\'\i}guez}, J.
  2006, \apj, 652, 1768

\bibitem[{{Blum} \& {Wurm}(2000)}]{BlumWurm2000}
{Blum}, J. \& {Wurm}, G. 2000, \icarus, 143, 138

\bibitem[{{Blum} \& {Wurm}(2008)}]{blum2008}
{Blum}, J. \& {Wurm}, G. 2008, \araa, 46, 21

\bibitem[{{Chiang} \& {Youdin}(2010)}]{ChiangYoudin2010}
{Chiang}, E. \& {Youdin}, A.~N. 2010, Annual Review of Earth and Planetary
  Sciences, 38, 493

\bibitem[{{Colwell} {et~al.}(2008){Colwell}, {Sture}, {Cintala}, {Durda},
  {Hendrix}, {Goudie}, {Curtis}, {Ashcom}, {Kanter}, {Keohane}, {Lemos},
  {Lupton}, \& {Route}}]{Colwell2008}
{Colwell}, J.~E., {Sture}, S., {Cintala}, M., {et~al.} 2008, \icarus, 195, 908

\bibitem[{{Cuzzi} {et~al.}(2003){Cuzzi}, {Davis}, \&
  {Dobrovolskis}}]{cuzzi2003}
{Cuzzi}, J.~N., {Davis}, S.~S., \& {Dobrovolskis}, A.~R. 2003, \icarus, 166,
  385

\bibitem[{{de Beule} {et~al.}(2013){de Beule}, {Kelling}, {Wurm}, {Teiser}, \&
  {Jankowski}}]{debeule2013}
{de Beule}, C., {Kelling}, T., {Wurm}, G., {Teiser}, J., \& {Jankowski}, T.
  2013, \apj, 763, 11

\bibitem[{{Deckers} \& {Teiser}(2013)}]{Deckers2013}
{Deckers}, J. \& {Teiser}, J. 2013, ArXiv e-prints

\bibitem[{{Dittrich} {et~al.}(2013){Dittrich}, {Klahr}, \&
  {Johansen}}]{dittrich2013}
{Dittrich}, K., {Klahr}, H., \& {Johansen}, A. 2013, \apj, 763, 117

\bibitem[{{Dominik} \& {Tielens}(1997)}]{DominikTielens1997}
{Dominik}, C. \& {Tielens}, A.~G.~G.~M. 1997, \apj, 480, 647

\bibitem[{{Dullemond} \& {Dominik}(2005)}]{Dominik2005}
{Dullemond}, C.~P. \& {Dominik}, C. 2005, \aap, 434, 971

\bibitem[{{Garaud} {et~al.}(2013){Garaud}, {Meru}, {Galvagni}, \&
  {Olczak}}]{garaud2013}
{Garaud}, P., {Meru}, F., {Galvagni}, M., \& {Olczak}, C. 2013, \apj, 764, 146

\bibitem[{{Geretshauser} {et~al.}(2011){Geretshauser}, {Meru}, {Speith}, \&
  {Kley}}]{GeretshauserEtal2011}
{Geretshauser}, R.~J., {Meru}, F., {Speith}, R., \& {Kley}, W. 2011, \aap, 531,
  A166

\bibitem[{{Goldreich} \& {Ward}(1973)}]{GoldreichWard1973}
{Goldreich}, P. \& {Ward}, W.~R. 1973, \apj, 183, 1051

\bibitem[{{G{\"u}ttler} {et~al.}(2009){G{\"u}ttler}, {Krause}, {Geretshauser},
  {Speith}, \& {Blum}}]{GuettlerEtal2009}
{G{\"u}ttler}, C., {Krause}, M., {Geretshauser}, R.~J., {Speith}, R., \&
  {Blum}, J. 2009, \apj, 701, 130

\bibitem[{{Jankowski} {et~al.}(2012){Jankowski}, {Wurm}, {Kelling}, {Teiser},
  {Sabolo}, {Guti{\'e}rrez}, \& {Bertini}}]{JankowskiEtal2012}
{Jankowski}, T., {Wurm}, G., {Kelling}, T., {et~al.} 2012, \aap, 542, A80

\bibitem[{{Johansen} {et~al.}(2007){Johansen}, {Oishi}, {Mac Low}, {Klahr},
  {Henning}, \& {Youdin}}]{JohansenEtal2007}
{Johansen}, A., {Oishi}, J.~S., {Mac Low}, M.-M., {et~al.} 2007, \nat, 448,
  1022

\bibitem[{{Kataoka} {et~al.}(2013){Kataoka}, {Tanaka}, {Okuzumi}, \&
  {Wada}}]{Kataoka2013}
{Kataoka}, A., {Tanaka}, H., {Okuzumi}, S., \& {Wada}, K. 2013, \aap, 554, A4

\bibitem[{{Kelling} \& {Wurm}(2009)}]{kelling2009}
{Kelling}, T. \& {Wurm}, G. 2009, \prl, 103, 215502

\bibitem[{{Kelling} {et~al.}(2013){Kelling}, {Wurm}, \&
  {K{\"o}ster}}]{Kelling2013}
{Kelling}, T., {Wurm}, G., \& {K{\"o}ster}, M. 2013, \apj, submitted

\bibitem[{{Klahr} \& {Lin}(2005)}]{Klahr2005}
{Klahr}, H. \& {Lin}, D.~N.~C. 2005, \apj, 632, 1113

\bibitem[{{Kocifaj} {et~al.}(2010){Kocifaj}, {Kla{\v c}ka}, {Wurm}, {Kelling},
  \& {Koh{\'u}t}}]{kocifaj2010}
{Kocifaj}, M., {Kla{\v c}ka}, J., {Wurm}, G., {Kelling}, T., \& {Koh{\'u}t}, I.
  2010, \mnras, 404, 1512

\bibitem[{{Kothe} {et~al.}(2010){Kothe}, {G{\"u}ttler}, \& {Blum}}]{Kothe2010}
{Kothe}, S., {G{\"u}ttler}, C., \& {Blum}, J. 2010, \apj, 725, 1242

\bibitem[{Krause {et~al.}(2011)Krause, Blum, Skorov, \& Trieloff}]{krause2011}
Krause, M., Blum, J., Skorov, Y., \& Trieloff, M. 2011, \icarus, {Im Druck}

\bibitem[{{Langkowski} {et~al.}(2008){Langkowski}, {Teiser}, \&
  {Blum}}]{langkowski2008}
{Langkowski}, D., {Teiser}, J., \& {Blum}, J. 2008, \apj, 675, 764

\bibitem[{{Meakin} \& {Donn}(1988)}]{Meakin1988}
{Meakin}, P. \& {Donn}, B. 1988, \apjl, 329, L39

\bibitem[{{Meisner} {et~al.}(2012){Meisner}, {Wurm}, \& {Teiser}}]{meisner2012}
{Meisner}, T., {Wurm}, G., \& {Teiser}, J. 2012, \aap, 544, A138

\bibitem[{{Meru} {et~al.}(2013){Meru}, {Geretshauser}, {Schaefer}, {Speith}, \&
  {Kley}}]{Meru2013}
{Meru}, F., {Geretshauser}, R.~J., {Schaefer}, C., {Speith}, R., \& {Kley}, W.
  2013, ArXiv e-prints

\bibitem[{{Okuzumi}(2009)}]{Okuzumi2009}
{Okuzumi}, S. 2009, \apj, 698, 1122

\bibitem[{{Okuzumi} {et~al.}(2012){Okuzumi}, {Tanaka}, {Kobayashi}, \&
  {Wada}}]{Okuzumi2012}
{Okuzumi}, S., {Tanaka}, H., {Kobayashi}, H., \& {Wada}, K. 2012, \apj, 752,
  106

\bibitem[{{Olofsson} {et~al.}(2009){Olofsson}, {Augereau}, {van Dishoeck},
  {Mer{\'{\i}}n}, {Lahuis}, {Kessler-Silacci}, {Dullemond}, {Oliveira},
  {Blake}, {Boogert}, {Brown}, {Evans}, {Geers}, {Knez}, {Monin}, \&
  {Pontoppidan}}]{olofsson2009}
{Olofsson}, J., {Augereau}, J.-C., {van Dishoeck}, E.~F., {et~al.} 2009, \aap,
  507, 327

\bibitem[{{Ormel} {et~al.}(2007){Ormel}, {Spaans}, \& {Tielens}}]{Ormel2007}
{Ormel}, C.~W., {Spaans}, M., \& {Tielens}, A.~G.~G.~M. 2007, \aap, 461, 215

\bibitem[{{Ossenkopf}(1993)}]{Ossenkopf1993}
{Ossenkopf}, V. 1993, \aap, 280, 617

\bibitem[{{Paraskov} {et~al.}(2006){Paraskov}, {Wurm}, \&
  {Krauss}}]{paraskov2006}
{Paraskov}, G.~B., {Wurm}, G., \& {Krauss}, O. 2006, \apj, 648, 1219

\bibitem[{{Paszun} \& {Dominik}(2009)}]{PaszunDominik2009}
{Paszun}, D. \& {Dominik}, C. 2009, \aap, 507, 1023

\bibitem[{{Poppe} {et~al.}(2010){Poppe}, {G{\"u}ttler}, \&
  {Springborn}}]{Poppe2010}
{Poppe}, T., {G{\"u}ttler}, C., \& {Springborn}, T. 2010, Earth, Planets, and
  Space, 62, 53

\bibitem[{{Ros} \& {Johansen}(2013)}]{RosJohansen2013}
{Ros}, K. \& {Johansen}, A. 2013, \aap, 552, A137

\bibitem[{Safronov(1969)}]{Safronov1969}
Safronov, V.~S. 1969, Evolution of the Protoplanetary Cloud and Formation of
  the Earth and the Planets, Moscow: Nauka Press, trans. NASA TTF 677, 1972

\bibitem[{{Saito} \& {Sirono}(2011)}]{Saito2011}
{Saito}, E. \& {Sirono}, S.-i. 2011, \apj, 728, 20

\bibitem[{{Sch{\"a}fer} {et~al.}(2007){Sch{\"a}fer}, {Speith}, \&
  {Kley}}]{SchaeferEtal2007}
{Sch{\"a}fer}, C., {Speith}, R., \& {Kley}, W. 2007, \aap, 470, 733

\bibitem[{{Schr{\"a}pler} \& {Blum}(2011)}]{Schraepler2011}
{Schr{\"a}pler}, R. \& {Blum}, J. 2011, \apj, 734, 108

\bibitem[{{Schr{\"a}pler} {et~al.}(2012){Schr{\"a}pler}, {Blum}, {Seizinger},
  \& {Kley}}]{SchraeplerEtal2012}
{Schr{\"a}pler}, R., {Blum}, J., {Seizinger}, A., \& {Kley}, W. 2012, \apj,
  758, 35

\bibitem[{{Seizinger} {et~al.}(2012){Seizinger}, {Speith}, \&
  {Kley}}]{Seizinger2012}
{Seizinger}, A., {Speith}, R., \& {Kley}, W. 2012, \aap, 541, A59

\bibitem[{{Sirono} {et~al.}(2006){Sirono}, {Satomi}, \&
  {Watanabe}}]{Sirono2006}
{Sirono}, S., {Satomi}, K., \& {Watanabe}, S. 2006, Journal of Geophysical
  Research (Solid Earth), 111, 6309

\bibitem[{{Sirono}(2011)}]{Sirono2011}
{Sirono}, S.-i. 2011, \apj, 735, 131

\bibitem[{{Suyama} {et~al.}(2008){Suyama}, {Wada}, \& {Tanaka}}]{Suyama2008}
{Suyama}, T., {Wada}, K., \& {Tanaka}, H. 2008, \apj, 684, 1310

\bibitem[{{Tanaka} {et~al.}(2013){Tanaka}, {Yamamoto}, {Tanaka}, {Miura},
  {Nagasawa}, \& {Nakamoto}}]{TanakaEtal2013}
{Tanaka}, K.~K., {Yamamoto}, T., {Tanaka}, H., {et~al.} 2013, \apj, 764, 120

\bibitem[{{Teiser} {et~al.}(2011{\natexlab{a}}){Teiser}, {Engelhardt}, \&
  {Wurm}}]{TeiserEtal2011b}
{Teiser}, J., {Engelhardt}, I., \& {Wurm}, G. 2011{\natexlab{a}}, \apj, 742, 5

\bibitem[{{Teiser} {et~al.}(2011{\natexlab{b}}){Teiser}, {K{\"u}pper}, \&
  {Wurm}}]{TeiserEtal2011a}
{Teiser}, J., {K{\"u}pper}, M., \& {Wurm}, G. 2011{\natexlab{b}}, \icarus, 215,
  596

\bibitem[{{Teiser} \& {Wurm}(2009{\natexlab{a}})}]{TeiserWurm2009a}
{Teiser}, J. \& {Wurm}, G. 2009{\natexlab{a}}, \aap, 505, 351

\bibitem[{{Teiser} \& {Wurm}(2009{\natexlab{b}})}]{TeiserWurm2009b}
{Teiser}, J. \& {Wurm}, G. 2009{\natexlab{b}}, \mnras, 393, 1584

\bibitem[{{van Boekel} {et~al.}(2005){van Boekel}, {Min}, {Waters}, {de Koter},
  {Dominik}, {van den Ancker}, \& {Bouwman}}]{vanBoekel2005}
{van Boekel}, R., {Min}, M., {Waters}, L.~B.~F.~M., {et~al.} 2005, \aap, 437,
  189

\bibitem[{{Wada} {et~al.}(2009){Wada}, {Tanaka}, {Suyama}, {Kimura}, \&
  {Yamamoto}}]{WadaEtal2009}
{Wada}, K., {Tanaka}, H., {Suyama}, T., {Kimura}, H., \& {Yamamoto}, T. 2009,
  \apj, 702, 1490

\bibitem[{{Wada} {et~al.}(2011){Wada}, {Tanaka}, {Suyama}, {Kimura}, \&
  {Yamamoto}}]{WadaEtal2011}
{Wada}, K., {Tanaka}, H., {Suyama}, T., {Kimura}, H., \& {Yamamoto}, T. 2011,
  \apj, 737, 36

\bibitem[{{Weidenschilling} {et~al.}(1989){Weidenschilling}, {Donn}, \&
  {Meakin}}]{Weidenschilling1989}
{Weidenschilling}, S.~J., {Donn}, B.~D., \& {Meakin}, P. 1989, in The Formation
  and Evolution of Planetary Systems, ed. H.~A. {Weaver} \& L.~{Danly},
  131--146

\bibitem[{{Weidling} {et~al.}(2009){Weidling}, {G{\"u}ttler}, {Blum}, \&
  {Brauer}}]{WeidlingEtal2009}
{Weidling}, R., {G{\"u}ttler}, C., {Blum}, J., \& {Brauer}, F. 2009, \apj, 696,
  2036

\bibitem[{{Windmark} {et~al.}(2012{\natexlab{a}}){Windmark}, {Birnstiel},
  {G{\"u}ttler}, {Blum}, {Dullemond}, \& {Henning}}]{windmark2012}
{Windmark}, F., {Birnstiel}, T., {G{\"u}ttler}, C., {et~al.}
  2012{\natexlab{a}}, \aap, 540, A73

\bibitem[{{Windmark} {et~al.}(2012{\natexlab{b}}){Windmark}, {Birnstiel},
  {Ormel}, \& {Dullemond}}]{Windmark2012b}
{Windmark}, F., {Birnstiel}, T., {Ormel}, C.~W., \& {Dullemond}, C.~P.
  2012{\natexlab{b}}, \aap, 544, L16

\bibitem[{{Wurm} \& {Blum}(1998)}]{Wurm1998}
{Wurm}, G. \& {Blum}, J. 1998, \icarus, 132, 125

\bibitem[{{Wurm} \& {Krauss}(2006)}]{Wurm2006b}
{Wurm}, G. \& {Krauss}, O. 2006, Physical Review Letters, 96, 134301

\bibitem[{{Wurm} {et~al.}(2005){Wurm}, {Paraskov}, \& {Krauss}}]{wurm2005}
{Wurm}, G., {Paraskov}, G., \& {Krauss}, O. 2005, \icarus, 178, 253

\bibitem[{{Youdin} \& {Johansen}(2007)}]{YoudinJohansen2007}
{Youdin}, A. \& {Johansen}, A. 2007, \apj, 662, 613

\bibitem[{{Zsom} {et~al.}(2010){Zsom}, {Ormel}, {G{\"u}ttler}, {Blum}, \&
  {Dullemond}}]{ZsomEtal2010}
{Zsom}, A., {Ormel}, C.~W., {G{\"u}ttler}, C., {Blum}, J., \& {Dullemond},
  C.~P. 2010, \aap, 513, A57

\end{thebibliography}

\end{document}